\documentclass[trackchanges,twocolumn]{aastex701}
\usepackage{graphicx}	
\usepackage{amsmath}	

\usepackage[T1]{fontenc}

\usepackage{CJKutf8}

\newcommand{\St}{\tau_{s}}

\newcommand{\vect}[1]{\boldsymbol{#1}}

\newcommand{\unitdir}[1]{\hat{\boldsymbol{#1}}}
\DeclareMathOperator{\arctantwo}{arctan2}

\graphicspath{{./}{figures/}}

\begin{document}

\title{Azimuthal Dust Polarization from Aerodynamically Aligned Grains as Evidence for the Streaming Instability in Protoplanetary Disks}

\author[orcid=0000-0001-7233-4171]{Zhe-Yu Daniel Lin (\begin{CJK*}{UTF8}{bkai}林哲宇\end{CJK*})}
\altaffiliation{Jansky Fellow of the National Radio Astronomy Observatory}
\affiliation{National Radio Astronomy Observatory, 520 Edgemont Road, Charlottesville, VA 22903, USA
}
\email[show]{dlin@nrao.edu}

\author[orcid=0000-0003-2719-6640,gname=Jeonghoon,sname=Lim]{Jeonghoon Lim (\begin{CJK}{UTF8}{mj}임정훈\end{CJK})}
\affiliation{Department of Physics and Astronomy, Iowa State University, Ames, IA 50010, USA}
\email{jhlim@iastate.edu}

\author[orcid=0000-0002-3771-8054]{Jacob B. Simon}
\affiliation{Department of Physics and Astronomy, Iowa State University, Ames, IA 50010, USA}
\email{jbsimon@iastate.edu}

\author[orcid=0000-0002-7402-6487]{Zhi-Yun Li}
\affiliation{Department of Astronomy, University of Virginia, 530 McCormick Rd.,
Charlottesville 22904, USA}
\email{zl4h@virginia.edu}

\author[orcid=0000-0001-6259-3575]{Daniel Carrera}
\affiliation{New Mexico State University, Department of Astronomy, PO Box 30001 MSC 4500, Las Cruces, NM 88001, USA}
\email{carrera4@nmsu.edu}

\author[orcid=0000-0001-5811-0454]{Manuel Fern\'andez-L\'opez}
\affiliation{Institut de Ciències de l'Espai (ICE), CSIC, Campus UAB, Carrer de Can Magrans s/n, 08193 Bellaterra, Barcelona, Spain}
\affiliation{Instituto Argentino de Radioastronomía (CCT-La Plata, CONICET; UNLP; CICPBA), C.C. No. 5, 1894, Villa Elisa, Buenos Aires, Argentina}
\affiliation{Facultad de Ciencias Astronómicas y Geofísicas, Universidad Nacional de La Plata, Paseo del Bosque S/N, B1900FWA La Plata, Argentina}
\email{manferna@gmail.com}

\author[orcid=0000-0003-2118-4999]{Rachel Harrison}
\affiliation{Department of Physics and Astronomy, Rice University, 6100 Main Street—MS 108, Houston, TX 77005, USA}
\email{rh128@rice.edu}

\author[orcid=0000-0001-9222-4367]{Rixin Li (\begin{CJK*}{UTF8}{bkai}李日新\end{CJK*})}
\altaffiliation{51 Pegasi b Fellow}
\affiliation{Department of Astronomy, University of California, Berkeley, CA 94720, USA}
\email{rixin@berkeley.edu}

\author[orcid=0000-0002-4540-6587]{Leslie W. Looney}
\affiliation{Department of Astronomy, University of Illinois, 1002 W Green St., Urbana, IL 61801, USA}
\email{lwl@illinois.edu}

\author[orcid=0000-0003-3017-4418]{Ian W. Stephens}
\affiliation{Department of Earth, Environment, and Physics, Worcester State University, Worcester, MA 01602, USA}
\email{istephens@worcester.edu}

\author[orcid=0000-0002-8537-6669]{Haifeng Yang}
\affiliation{Institute for Astronomy, School of Physics, Zhejiang University, 886 Yuhangtang Road, Hangzhou, 310027 Zhejiang, People’s Republic of China}
\affiliation{Center for Cosmology and Computational Astrophysics, Institute for Advanced Study in Physics, Zhejiang University, Hangzhou 310027, People's Republic of China}
\email{hfyangpku@gmail.com}

\begin{abstract}

(Sub)millimeter dust polarization in protoplanetary disks has revealed the presence of large ($\sim 100$~$\mu$m) dust grains that are aligned along their long axis following the azimuthal direction of the disk. The novel badminton birdie-like aerodynamic alignment predicts large grains to align with their long axes following the direction of gas flow experienced by the dust, denoted as the $\vect{A}$-field. With 3D streaming instability (SI) simulations, we find that the $\vect{A}$-field is predominantly in the radial direction in regions of low dust-to-gas ratio, but in the azimuthal direction in regions of high dust-to-gas ratio. Through polarized radiation transfer, we find that the resulting polarization angle indeed follows the disk azimuthal direction in the high dust density regions. Therefore, ongoing SI is an attractive candidate for producing the azimuthal dust polarization pattern as observed in an increasing number of disks.

\end{abstract}



\section{Introduction} \label{sec:intro}

Dust grains play a crucial role in protoplanetary disks, serving as the raw material for planetesimals and planets, and also determining the structure and evolution of disks \citep[e.g.][]{Drazkowska2023ASPC..534..717D, Birnstiel2024ARA&A..62..157B}. 
Spatially resolved millimeter-wave polarization has provided critical insights into the properties of dust in these disks \citep[e.g.][]{Kataoka2016ApJ...831L..12K, Stephens2017ApJ...851...55S, Lee2018ApJ...854...56L, Girart2018ApJ...856L..27G, Ohashi2018ApJ...864...81O, Alves2018A&A...616A..56A, Sadavoy2018ApJ...869..115S, Bacciotti2018ApJ...865L..12B, Dent2019MNRAS.482L..29D, Takahashi2019ApJ...872...70T, Harrison2019ApJ...877L...2H, Sadavoy2019ApJS..245....2S, Stephens2020ApJ...901...71S, Aso2021ApJ...920...71A, Hull2022ApJ...930...49H, Tang2023ApJ...947L...5T, Stephens2023Natur.623..705S, Lin2024MNRAS.528..843L, Liu2024ApJ...963..104L, Harrison2024ApJ...967...40H, Looney2025ApJ...985..148L, Cortes2025ApJ...992L..31C}. 
Since the first polarization detections at disk scale \citep{Rao2014ApJ...780L...6R, Stephens2014Natur.514..597S}, it is now abundantly clear that the origin of polarization can be due to scattering \citep[e.g.][]{Kataoka2015ApJ...809...78K, Yang2016MNRAS.456.2794Y, Yang2017MNRAS.472..373Y}, aligned grains \citep[e.g.][]{Alves2018A&A...616A..56A, Looney2025ApJ...985..148L, Fourkas2026ApJ...999....4F}, or both effects combined \citep[e.g.][]{Yang2016MNRAS.460.4109Y, Yang2019MNRAS.483.2371Y, Lin2022MNRAS.512.3922L, Stephens2023Natur.623..705S}.

The origin of aligned grains in protoplanetary disks is currently a mystery, but we list a few key empirical constraints on the geometrical properties of grains. First, the fact that grains can scatter millimeter-wavelength radiation requires grain sizes to be at least $\sim 100$~$\mu$m \citep[e.g.][]{Kataoka2015ApJ...809...78K, Yang2016MNRAS.456.2794Y, Lin2020MNRAS.496..169L}, which is significantly larger than the typical $\sim 0.1$~$\mu$m interstellar medium grains. Second, while grains are generally triaxial, polarization infers grains aligned predominantly by the \textit{long} axes making them effectively act like prolate grains for polarization. Third, the alignment of the grain's long axis is in the azimuthal direction around the disk. These constraints come simply from the axisymmetry of disks, which determines the azimuthal variation of polarization fraction and polarization angle \citep{Yang2019MNRAS.483.2371Y, Mori2021ApJ...908..153M}. 
An addition to the third constraint is that in some cases, there are reported deviations to alignment in the azimuthal direction by $4.5^{\circ}$ for AS 209 \citep{Mori2019ApJ...883...16M}, $7.3^{\circ}$ for GG Tau \citep{Tang2023ApJ...947L...5T}, and $4^{\circ}$ for HL~Tau (Y.-W. Xu et al. in preparation). Relative to the disk rotation directions, all are trailing spirals. 

Explaining how and why the large grains are azimuthally aligned and effectively prolate in shape has proven difficult. 
Historically, grains in the interstellar medium (ISM) are shown to be aligned to the magnetic field ($\vect{B}$-field) via RAdiative Torques (RAT) \citep[e.g.][]{Draine1996ApJ...470..551D, Lazarian2007MNRAS.378..910L, Andersson2015ARA&A..53..501A}. In short, as a grain spins, the axis of maximum inertia (i.e., the short axis) becomes aligned to its angular momentum vector \citep[e.g.][]{Purcell1979ApJ...231..404P} and both become aligned to the $\vect{B}$-field. Over an ensemble, which determines the polarization, the population appears as effectively oblate grains whose \textit{short} axes are aligned to $\vect{B}$-field. Applications of RAT to protoplanetary disk geometry with the expected toroidal $\vect{B}$-field structure produce radially oriented polarization \citep[e.g.][]{Cho2007ApJ...669.1085C, Yang2016MNRAS.460.4109Y}. 
The observed azimuthally aligned, effectively prolate grains immediately rule out the conventional picture applied to the ISM and call for new physics to the grain alignment problem of protoplanetary disks.

Recently, \cite{Lin2024MNRAS.534.3713L} proposed the badminton birdie-like aerodynamic alignment mechanism (hereafter birdie-like alignment), where an offset in the grain center of mass to the general geometric center along the long axis of large, millimeter-sized grains can produce the necessary effectively prolate grains, simply due to gas drag from gas motion relative to the grain\footnote{Other alignment scenarios are discussed in Section~\ref{sec:discussion}}. Dust polarization traces the alignment direction of the grain following this relative motion, which we refer to as the field of aerodynamic flow --- the $\vect{A}$-field --- defined as the gas velocity relative to the dust velocity. Grains are aligned of order the stopping time, which is usually a fraction of the Keplerian time for millimeter grains \citep[e.g.][]{Birnstiel2024ARA&A..62..157B}. The speed of alignment is a major advantage of this mechanism, especially in high-density disk environments where other mechanisms might be damped. 

Although birdie-like alignment can readily satisfy the polarization requirements on the shapes and sizes of grains, the observed alignment direction requires an $\vect{A}$-field that is in the \textit{azimuthal} direction, which is unexpected. 
Assuming that dust grains experience aerodynamic drag in a gaseous disk with the typical negative pressure gradient in steady state, the grains with Stokes number $\ll 1$ (such as the expected $100$~$\mu$m grains) should drift inward relative to the gas \citep{Takeuchi2002ApJ...581.1344T}. As a result, the $\vect{A}$-field should point in the outward radial direction, which would produce \textit{radially} oriented polarization under birdie-like alignment \citep{Lin2024MNRAS.534.3713L}. This is inconsistent with the observations of azimuthal polarization patterns, highlighting the tension between simple equilibrium dust-gas dynamics and the observations.

One potential avenue is to incorporate dust feedback where local gas dynamics within a concentrated dust layer can be significantly altered by dust drag \citep{Nakagawa1986Icar...67..375N}. 
Furthermore, the time-dependent evolution of the coupled gas and dust dynamics is regularly studied in simulations of Streaming Instability (SI; e.g., \citealt{Johansen2007ApJ...662..627J}). First studied by \cite{Youdin2005ApJ...620..459Y} in its linear regime, SI has long thought to be a promising avenue for forming planetesimals directly from relatively small grains, involving significant clumping of dust through collective aerodynamic effects \citep[e.g.][]{Johansen2007ApJ...662..627J, Chiang2010AREPS..38..493C}. 
Since SI necessarily involves aerodynamic interactions between dust and gas, it begs the question: how do the flows align grains through birdie-like alignment? 
As explained in \cite{Squire2020MNRAS.498.1239S}, inward gas motion is deflected to the azimuthal direction due to the Coriolis force, making it a likely candidate for explaining the observed inferred azimuthal alignment direction.
Recent advancements in SI simulations with greatly improved resolution and coverage in 3D make it feasible to address the above question \citep{Lim2024ApJ...969..130L, Lim2025ApJ...981..160L, Lim2025ApJ...993...12L, Lim2026ApJ..1000..156L}.

In this paper, we demonstrate that SI naturally modifies the local aerodynamic flow experienced by dust grains, enabling birdie-like alignment to produce azimuthally aligned, effectively prolate grains.
The paper is organized as follows: Sec.~\ref{sec:problem_setup} describes the adopted SI simulations and its polarized radiation transfer. Sec.~\ref{sec:results} presents the resulting polarization images. Sec.~\ref{sec:discussion} discusses the implications and further compares other grain alignment theories. Our results are summarized in Sec.~\ref{sec:summary}. 

\section{Problem Setup} \label{sec:problem_setup}

\subsection{Streaming Instability Simulation}

We analyze existing SI simulations from \cite{Lim2026ApJ..1000..156L} and only briefly describe the methodology and scope (see also \citealt{Lim2024ApJ...969..130L, Lim2025ApJ...981..160L}).
To solve for the coupled dynamics of gas and dust, the simulations used the \texttt{ATHENA} code \citep{Stone2008ApJS..178..137S} to model the isothermal, unmagnetized gas, while utilizing the Lagrangian particle module from \cite{Bai2010ApJS..190..297B} to numerically solve the equation of motion for dust particles. We do not assume external turbulence and ignore self-gravity of solid particles. 

The simulations adopt the standard local shearing box approximation, which models a corotating patch of a disk, with Keplerian frequency $\Omega$ \citep{Hawley1995ApJ...440..742H, Stone2010ApJS..189..142S}. The models are fully three-dimensional and the local Cartesian coordinates $(x,y,z)$ denote the radial, azimuthal, and vertical directions, respectively. The unit directions are denoted by $\unitdir{e}_{x}$, $\unitdir{e}_{y}$, and $\unitdir{e}_{z}$, respectively. 
Boundary conditions include the standard shear-periodic boundary conditions in the $x$ direction, purely periodic boundaries in the $y$ direction, and an outflow boundary condition in $z$ \citep{Li2018ApJ...862...14L}. 

The volume mass density of the gas and dust particles is $\rho_{g}$ and $\rho_{p}$, respectively. 
The gas is initially in vertical hydrostatic balance, having a vertical Gaussian density profile with scale height $H$ and midplane density $\rho_{g0}$. The surface density is $\Sigma_{g0} = \sqrt{2\pi}\rho_{g0} H$.
Similarly, particles are initially distributed with a Gaussian density profile with an initial scale height ($H_{p0}$) of $0.025H$ in the vertical direction. Thus, the initial surface density of particles is $\Sigma_{p0}=\sqrt{2\pi}\rho_{p0}H_{p0}$.
The initial velocity field of the gas and dust is determined by the Nakagawa-Sekiya-Hayashi (NSH) equilibrium solutions \citep{Nakagawa1986Icar...67..375N}. 

The abundance of dust relative to gas is parameterized by 
\begin{align}
    Z \equiv \frac{ \Sigma_{p0} }{ \Sigma_{g0} }
\end{align}
which is an input for each simulation. For convenience, we define the local dust-to-gas ratio with 
\begin{align}
    \epsilon \equiv \frac{ \rho_{p} }{ \rho_{g} }.
\end{align}
which varies in the spatial domain. 
The effect from the size of the particle is controlled by the Stokes number (i.e., the dimensionless stopping time), 
\begin{align}
    \St \equiv t_{\text{stop}} \Omega
\end{align}
where $t_{\text{stop}}$ is the stopping time of the particle and $\Omega$ is the Keplerian frequency.
We refer the reader to Table~1 of \cite{Lim2026ApJ..1000..156L} for the list of simulations with various $(\St,Z)$ and other details like the spatial size. 

The length, time, and mass units from the simulations are $H$, $\Omega^{-1}$, and $\rho_{g0} H^{3}$, respectively. The velocity unit is the isothermal sound speed $c_{s} \equiv H \Omega$. 
We obtain the gas velocity $\vect{v}_{g}$ and the dust velocity $\vect{v}_{p}$ as a function of $(x,y,z)$. 
The aerodynamic flow is defined as the gas velocity with respect to the center of mass of each grain 
\begin{align} \label{eq:A_definition}
    \vect{A} \equiv \vect{v}_{g} - \vect{v}_{p}
\end{align}
which has components $\vect{A} = A_{x} \unitdir{e}_{x} + A_{y} \unitdir{e}_{y} + A_{z} \unitdir{e}_{z}$.
The direction of $\vect{A}$ is a crucial quantity and can be characterized by the polar angle 
\begin{align} \label{eq:theta_from_A}
    \theta = \arccos \bigg(\frac{A_{z}}{|\vect{A}|} \bigg)
\end{align}
and the azimuthal direction is
\begin{align} \label{eq:phi_from_A}
    \phi = \arctantwo (A_{y}, A_{x})
\end{align}
where the function $\arctantwo$ computes the angle in the appropriate quadrant defined with respect to the $x$-axis following the right-hand rule. 

\subsection{Polarized Radiation Transfer}

As a demonstration, we assume the lines of sight are in the optically thin limit. 
Being able to identify polarization from aligned grains rather than from scattering in observations automatically means disks have optical depths that are less than of order unity \citep[e.g.][]{Yang2017MNRAS.472..373Y, Lin2022MNRAS.512.3922L}.
Generally, this is relevant to the longer wavelengths ($\lambda > \sim 3$~mm) where the dust opacity continually decreases \citep[e.g.][]{Harrison2019ApJ...877L...2H, Lin2024MNRAS.528..843L}. 
Assuming the optically thin limit avoids assumptions about the absolute value of the dust opacities since it is the polarization angle that we are interested in. We can also ignore contribution from scattering and dichroic extinction in this limit. 

We calculate the resulting polarization from the face-on view of the SI simulations. 
The Stokes frame is defined such that positive $Q$ is along the $x$-axis and negative $Q$ is along the $y$-axis, while positive $U$ is along the $x=y$ direction and negative $U$ is along $x=-y$. The polarization angle is 
\begin{align} \label{eq:chi}
    \chi \equiv \frac{1}{2} \arctantwo ( U, Q)
\end{align}
where $\chi=0^{\circ}$ means polarization (E-vectors) parallel to the (radial) $x$-axis, and $\chi=90^{\circ}$ means polarization parallel to the (azimuthal) $y$-axis. The polarized intensity is $P\equiv \sqrt{Q^{2} + U^{2}}$, while the polarization fraction is $p \equiv P / I$.

The radiation transfer equation at frequency $\nu$ of the Stokes parameters $(I,Q,U,V)$ in the optically thin limit is 
\begin{align} \label{eq:solution_to_rt}
    \frac{d}{dz}
    \begin{pmatrix}
        I \\
        Q \\
        U\\
        V
    \end{pmatrix}
    = \rho_{p} B_{\nu}(T) \begin{pmatrix}
        \kappa_{1} \\
        \kappa_{2} \\
        \kappa_{3} \\
        0
    \end{pmatrix} 
\end{align}
where $B_{\nu}(T)$ is the blackbody radiation at frequency $\nu$ with temperature $T$. The vector quantity $\vect{\kappa}=(\kappa_{1},\kappa_{2},\kappa_{3},0)$ represents the absorption cross section per unit mass of grains in the Stokes frame. The last element is $0$ since grains do not emit circular polarization. The solution to Eq.~\ref{eq:solution_to_rt} is a simple integration along $z$ and we ignore background contribution. 

To determine $\vect{\kappa}$, we assume the grains are prolates and are aligned to the $\vect{A}$-field according to birdie-like alignment. The alignment timescale is comparable to the stopping time, and the stopping time is less than the Keplerian time for small $\St$. In reality, each grain may not be perfectly aligned with $\vect{A}$ as the grain oscillates, but the polarization from the ensemble follows $\vect{A}$ with an efficiency that depends on the grain elongation and alignment efficiency. After all, the characteristic oscillation time is on timescales of minutes to hours, which is drastically shorter than the Keplerian time, for representative disk environments \citep{Lin2024MNRAS.534.3713L}. 

By assuming a prolate grain, the polarized thermal emission does not have a Stokes~$U$ component in the Stokes frame along the projected symmetric axis of the grain (i.e., the grain frame). We can simplify the absorption vector to 
\begin{align}
    \vect{\kappa} = 
    \begin{pmatrix}
        \kappa_{1}' \\
        \kappa_{2}' \cos 2 \psi \\
        \kappa_{2}' \sin 2 \psi \\
        0 
    \end{pmatrix}
\end{align}
where $\kappa_{1}'$ and $\kappa_{2}'$ are the only two independent parameters in the grain frame and $\psi$ is the angle of rotation from the grain frame to the lab frame (see, e.g., \citealt{Lin2024MNRAS.528..843L}). Furthermore, we adopt the dipole approximation for prolate grains, which has a convenient polarization form
\begin{align}
    p'(\theta_{g}) = \frac{p_{0} \sin^{2} \theta_{g}}{1 - p_{0} \cos^{2} \theta_{g}}
\end{align}
where $\theta_{g}$ is the angle between the axis of symmetry to the observer and $p_{0}$ is the intrinsic polarization, which is $p'$ seen at $\theta_{g}=90^{\circ}$ \citep{Lee1985ApJ...290..211L, Yang2016MNRAS.460.4109Y}. 
We treat $p_{0}$ as a parameter that captures grain elongation/porosity \citep[e.g.][]{Potapov2025A&ARv..33....6P} and alignment efficiency \citep[e.g.][]{Lee1985ApJ...290..211L}. We adopt $p_{0}=0.1$ motivated by empirical measurements after accounting for optical depth and projection effects \citep[e.g.][]{Stephens2023Natur.623..705S, Lin2024MNRAS.528..843L, Fourkas2026ApJ...999....4F}. 
As a result, $\kappa_{1}'$ and $\kappa_{2}'$ are related by $\kappa_{2}' = p' \kappa_{1}'$. 
The absolute values of $\kappa_{1}'$ and $\kappa_{2}'$ do not impact the polarization angles in the optically thin limit. 
In this work, we focus on a face-on view, and it is simple to find that $\psi=\phi$ and $\theta_{g}=\theta$. 

\section{Results} \label{sec:results}

In this section, we first explore two representative simulations with different $(\St,Z)$ at particular snapshots in detail that exhibit clear $\vect{A}$ in the azimuthal direction with relatively small $\St$. We then explore the time dependence and survey all the available 3D simulations in $(\St,Z)$ space.

\begin{figure*}
    \centering
    \includegraphics[width=\textwidth]{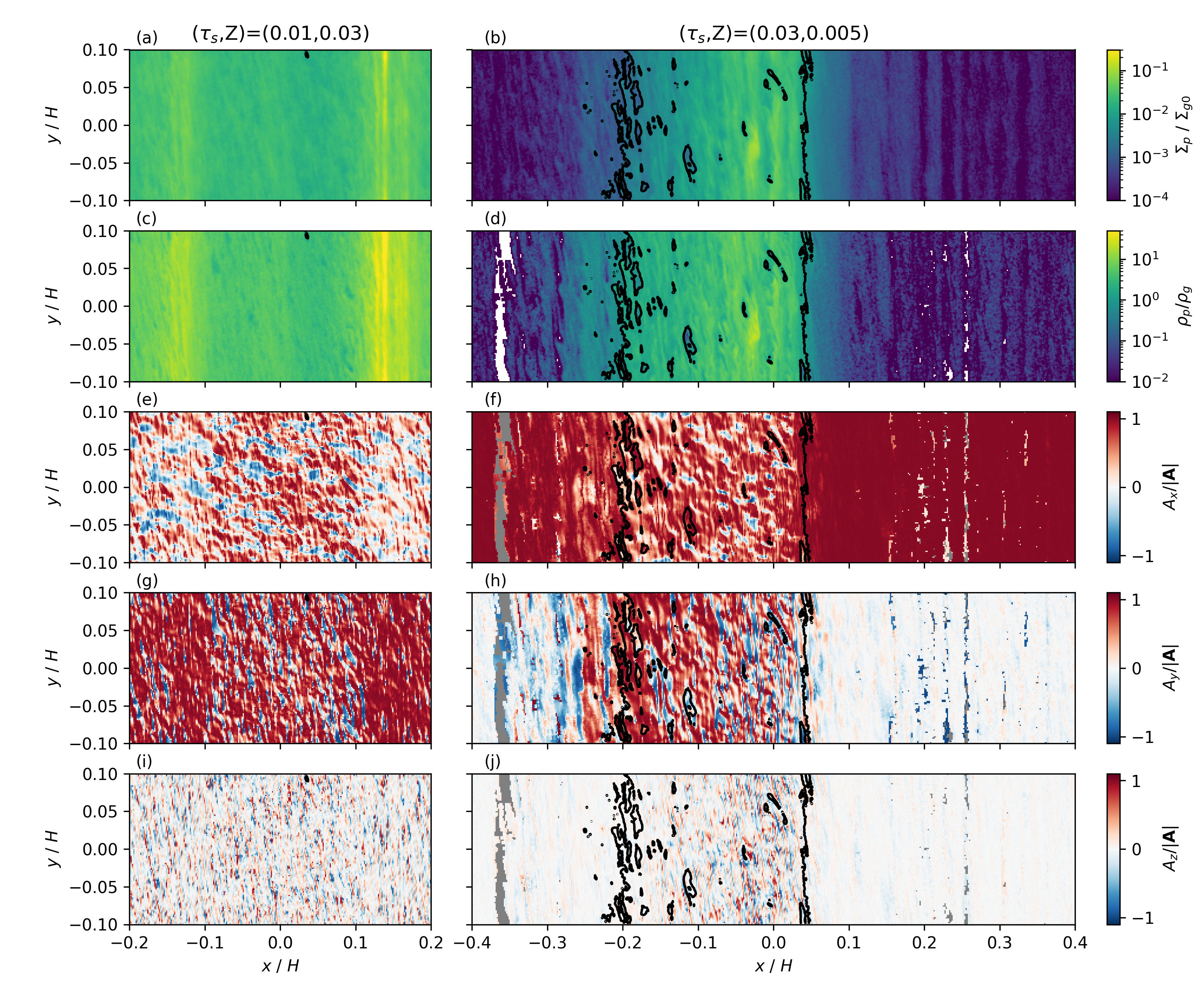}
    \caption{
        Structure of the simulations, where $x$ is the radial direction and $y$ is the azimuthal direction. The left and right columns correspond to simulations with $(\tau_{s},Z)=(0.01,0.03)$ and $(0.03,0.005)$. The top row is the dust surface density $\Sigma_{p}$, while the second row is the dust-to-gas ratio $\epsilon$ at the midplane. The third, fourth, and last rows correspond to the radial, azimuthal, and vertical components ($A_{x}$, $A_{y}$, and $A_{z}$) of the aerodynamic flow $\vect{A}$ at the midplane. 
        For reference, the median values of $|\vect{A}|$ at the midplane are $6 \times 10^{-4} c_{s}$ and $2.8 \times 10^{-3} c_{s}$, respectively. 
        In all panels, the black contours show where $\epsilon=1$. For $(\tau_{s},Z)=(0.01,0.03)$, the $\epsilon>1$ almost everywhere in the midplane. The gray regions for $(\tau_{s},Z)=(0.03,0.005)$ in panels f,h,j are where there is no dust present. 
        The two simulations demonstrate that $\vect{A}$ show strong positive, azimuthal components (tailwind) where $\epsilon>1$. 
    }
    \label{fig:midplane}
\end{figure*}

Fig.~\ref{fig:midplane} shows several quantities from two snapshots. Fig.~\ref{fig:midplane}a shows the surface density for the relatively large $Z$ (supersolar metalicity) case of $(\St, Z)=(0.01, 0.03)$ at $t\Omega=2450$, while Fig.~\ref{fig:midplane}c shows the dust-to-gas ratio $\epsilon$ at the midplane. At this $t$, $\epsilon$ is large with $\epsilon>1$ almost everywhere and can reach values as high as $\epsilon \sim 50$. Fig.~\ref{fig:midplane}e,g,i are the $x$, $y$, and $z$ components of $\vect{A}$ in the midplane, each normalized by the magnitude $|\vect{A}|$ to portray the relative strength. Clearly, in the midplane, $A_{z}/|\vect{A}|$ is negligible and thus $\vect{A}$ is predominantly determined by $A_{x}$ and $A_{y}$. Strikingly, Fig.~\ref{fig:midplane}e,g show that the azimuthal component is what dominates throughout the midplane and $A_{y}$ is generally positive which corresponds to a tailwind.\footnote{Since a dust grain is predominantly orbiting around the star ($\vect{v}_{p}$ is predominantly parallel to $\unitdir{e}_{y}$), a positive $A_{y}$ means the dust grain experiences a gaseous flow that is blowing in the direction of travel of the grain, which we describe as a ``tailwind" (or, in a stricter sense, $\vect{A}$ is parallel to $\vect{v}_{p}$). When the grain experiences a gaseous flow that is blowing opposite to the direction of travel of the grain, we describe $\vect{A}$ as a ``headwind" (or, in a stricter sense, $\vect{A}$ is antiparallel to $\vect{v}_{p}$). 
}
The tailwind likely matches the explanation from \cite{Squire2020MNRAS.498.1239S} where the inward gas motion is deflected into $+y$ azimuthal direction due to the Coriolis force. 
In contrast, a few patches in Fig.~\ref{fig:midplane}g exhibit strong negative $A_{y}$, which corresponds to a headwind.

The simulation with a relatively low $Z$ (subsolar metalicity), but larger grain case of $(\St,Z)=(0.03, 0.005)$ taken at $t\Omega=2500$ shows a distinctly different structure. 
Fig.~\ref{fig:midplane}b shows that $\Sigma_{p}$ is more confined ($-0.2 < x / H < 0.05$), which we call a filament, and the midplane $\epsilon$ varies significantly across the radial domain, generally following the trend from $\Sigma_{p}$. The midplane $\epsilon$ only reaches up to $\sim 10$, which is lower than in the previous example. Outside the filament, $\epsilon$ can be less than $0.01$. 

Fig.~\ref{fig:midplane}f,h, and j show the components of $\vect{A}$ and also show the contour where $\epsilon=1$.  The vertical component remains the smallest component (Fig.~\ref{fig:midplane}j) which we can effectively ignore. 
In regions where $\epsilon<1$, $\vect{A}$ is predominantly in the outward radial direction (Fig.~\ref{fig:midplane}f). This is expected from the equilibrium solutions, where small grains undergo radial drift when $\epsilon$ is low and thus the aerodynamic flow that the grain feels is in the outward radial direction. However, in regions where $\epsilon > 1$, $A_{x} / |\vect{A}|$ quickly drops and we see large nonzero values of $A_{y} / | \vect{A} |$ corresponding to azimuthal aerodynamic flow. Most of those regions exhibit strong positive $A_{y}$, particularly at the inner part of the filament ($x\sim -0.2 H$), while a few patches in the filament show strong negative $A_{y}$.

\begin{figure*}
    \centering
    \includegraphics[width=\textwidth]{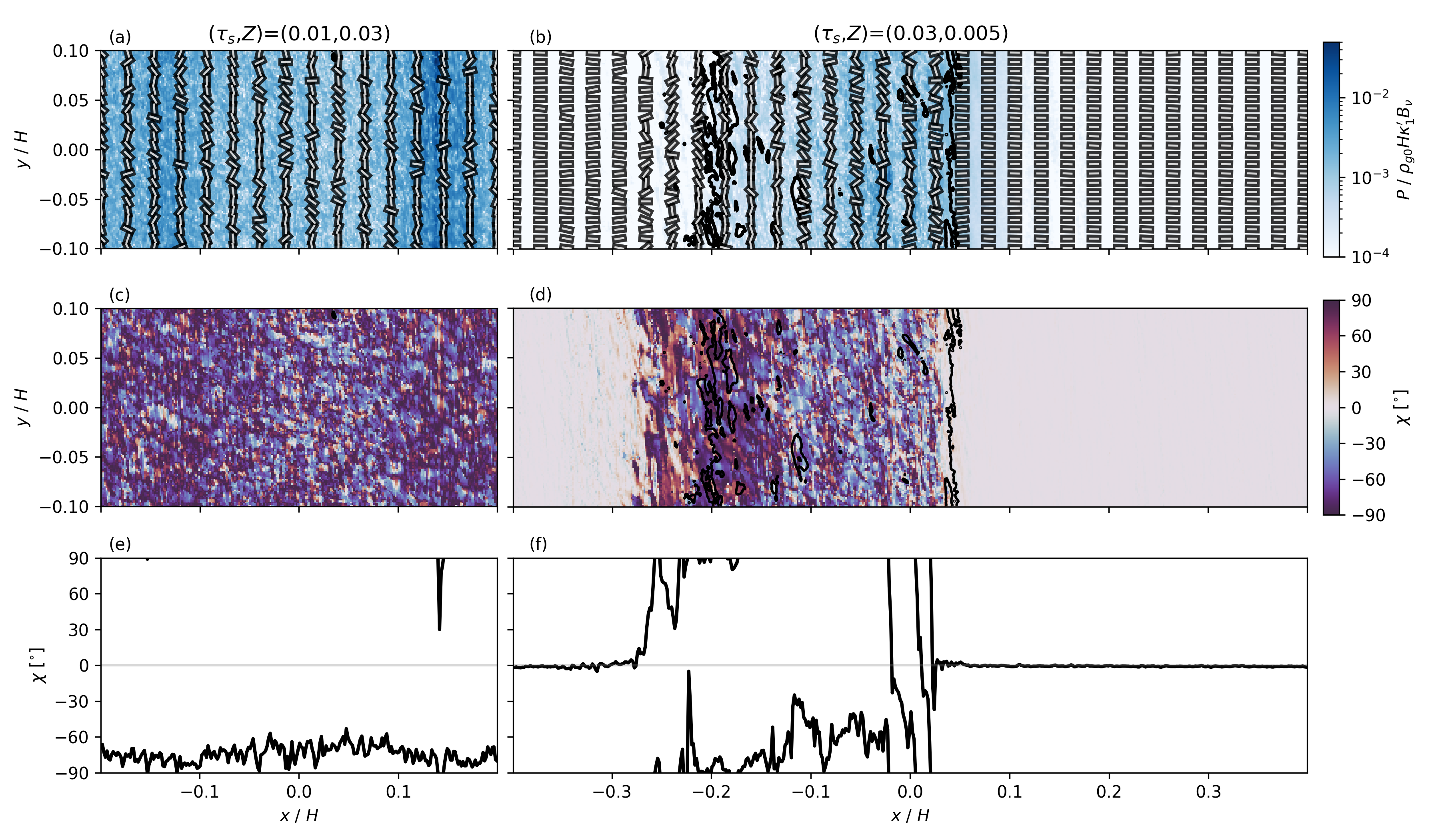}
    \caption{
        The polarization images for the two simulations. The left and right columns correspond to simulations with $(\tau_{s},Z)=(0.01,0.03)$ and $(0.03,0.005)$. The top row shows the polarized intensity with overplotted vectors denoting the polarization angles $\chi$. The second row shows the map of $\chi$ where $\chi=0^{\circ}$ is in the radial direction, and $\chi=\pm90^{\circ}$ is in the azimuthal direction The color map is cyclic, i.e., $\chi=90^{\circ}$ is equivalent to $\chi=-90^{\circ}$.
        The black contours for both rows mark where $\epsilon=1$. 
        The bottom row shows the azimuthally averaged $\chi$ by averaging Stokes~$Q$ and $U$ in the $y$-axis, then utilizing Eq.~\ref{eq:chi}. 
        Values of $\chi \in (0^{\circ}, 90^{\circ})$ are leading spirals, while $\chi \in (-90^{\circ}, 0^{\circ})$ are trailing spirals. 
    }
    \label{fig:pol2d}
\end{figure*}

Fig.~\ref{fig:pol2d} shows the resulting polarization images. For $(\St,Z)=(0.01,0.03)$, Fig.~\ref{fig:pol2d}a shows that the polarized intensity generally corresponds to the regions of higher $\Sigma_{p}$ (Fig.~\ref{fig:midplane}a) and that the polarization in those regions is predominantly in the azimuthal direction as expected from the dominating $A_{y}$ component (Fig.~\ref{fig:midplane}e,g). 
Fig.~\ref{fig:pol2d}c shows the complete view of $\chi$ without the effects of sampling necessary for plotting vectors and shows that $\chi$ is predominantly near $\chi \sim -90^{\circ}$. 

To better visualize the quantitative behavior of $\chi$, we average the Stokes parameters along the $y$-axis resulting in a 1D profile, denoted by $\langle I \rangle_{y}$, $\langle Q \rangle_{y}$, and $\langle U \rangle_{y}$.\footnote{The averaging is reasonable since $H$ is generally much smaller than the radius $R$ for a disk and the range in $y$ is only $0.2H$ making the patch especially small in the azimuthal direction and easily within a single observing beam.} Utilizing Eq.~\ref{eq:chi} with $\langle Q \rangle_{y}$ and $\langle U \rangle_{y}$ gives the azimuthally averaged $\chi$.
Fig.~\ref{fig:pol2d}e shows that the azimuthally averaged $\chi$ is $\sim -75^{\circ}$ and ranges from $-60^{\circ}$ to $-90^{\circ}$. 
Even though $A_{y}/|\vect{A}|$ can be negative (headwind) in certain patches (Fig.~\ref{fig:midplane}g), the polarization direction in Fig.~\ref{fig:pol2d}c is the same as other regions where $A_{y}/|\vect{A}|$ is positive (tailwind). This is because polarization is degenerate by $180^{\circ}$ and thus neighboring antiparallel $\vect{A}$ do not lead to cancellation of polarization under spatial averaging.

The value of $\chi \sim -75^{\circ}$ means a deviation from the prefect azimuthal direction corresponding to a trailing spiral, which is intriguing. From Fig.~\ref{fig:midplane}e and g, while positive $A_{y}$ is the dominating component, $A_{x}$ is also slightly positive, which means that the midplane $\vect{A}$ is a leading spiral. The resulting trailing spiral $\chi$ is due to variations of $\vect{A}$ in the vertical direction, which we explain in Appendix~\ref{sec:vertical_structure_of_A}. In short, $\vect{A}$ is a trailing spiral in regions away from the midplane, leading to some degree of cancellation along the line of sight.

For $(\St,Z)=(0.03,0.005)$, Fig.~\ref{fig:pol2d}b and d are the resulting $P$ and $\chi$, respectively. Given the large radial variation in $\Sigma_{p}$ (Fig.~\ref{fig:midplane}b), Fig.~\ref{fig:pol2d}d shows that high polarized intensity regions correspond to high midplane $\epsilon$ and that the polarization is predominantly oriented azimuthally in those regions. The regions of low polarized intensity have radially oriented polarization, corresponding to where the midplane $\epsilon < 1$. The more complete view of $\chi$ (Fig.~\ref{fig:pol2d}d) shows the contrast more clearly. 
Intriguingly, there is a slight radial gradient of $\chi$. Fig.~\ref{fig:pol2d}f shows the 1D profile of $\chi$, azimuthally averaged like in the previous case. At $x=-0.2H$, $\chi$ is on average $\sim -90^{\circ}$, but increases to $\sim -60^{\circ}$ at $x=-0.05 H$. 

Both simulations show $\chi \sim -90^{\circ}$ where $\epsilon$ is high with a slight preference for trailing spirals. As mentioned in Sec.~\ref{sec:intro}, a few observations have also shown evidence of trailing spirals with values of $\chi \sim -85^{\circ}$ when adopting our coordinate system \citep{Mori2019ApJ...883...16M, Tang2023ApJ...947L...5T}. While it is currently much too premature to quantitatively compare $\chi$ against observations without considering effects of finite angular resolution, the trailing spirals are at least consistent qualitatively.

\begin{figure*}
    \centering
    \includegraphics[width=\textwidth]{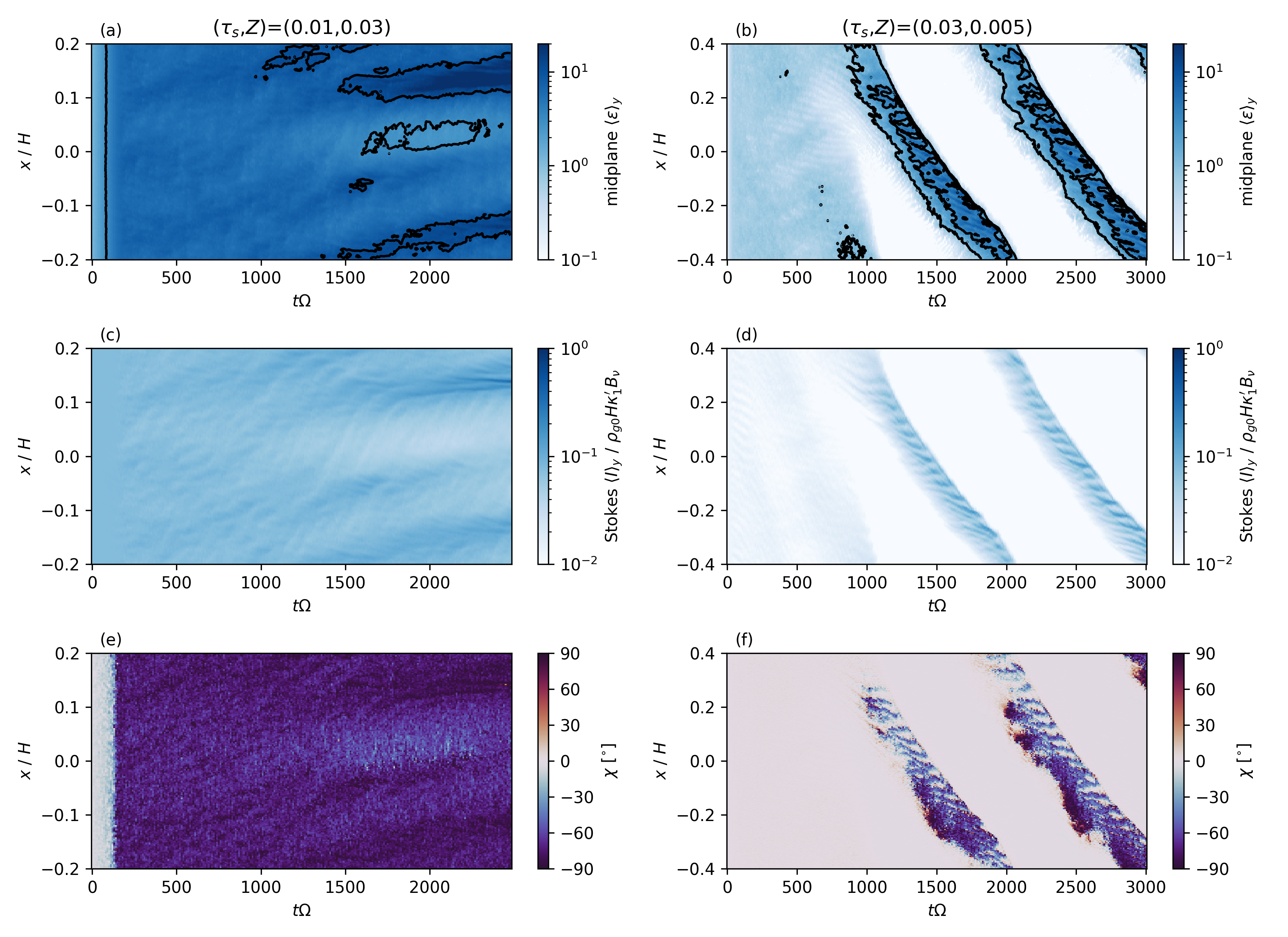}
    \caption{
        Top panels: the $x$-axis profile of the midplane $\epsilon$ averaged along the $y$-axis plotted against time (the horizontal axis). The contours show $\langle \epsilon \rangle_{y} = [1, 3, 10]$. 
        Middle panels: the $x$-axis profile of Stokes~$\langle I \rangle_{y}$, which is Stokes~$I$ averaged along the $y$-axis. Bottom panels: the corresponding $\chi$ also from Stokes~$Q$ and $U$ averaged along the $y$-axis. The left and right columns correspond to the two simulations with $(\tau_{s},Z)=(0.01,0.03)$ and $(0.03,0.005)$, respectively. Note that the $x$-axis has shear-periodic boundary condition. 
    }
    \label{fig:spacetime}
\end{figure*}

We next investigate the time dependence of the simulations. We define $\langle \epsilon \rangle_{y}$ as the $\epsilon$ averaged over the $y$-axis. Fig.~\ref{fig:spacetime} shows the midplane $\langle \epsilon \rangle_{y}$, Stokes~$\langle I \rangle_{y}$ and the azimuthally averaged $\chi$ profiles against time. 
Initially, $\chi$ begins at $0^{\circ}$, where dust is predominantly radially drifting inward relative to the gas for both cases.
For $(\St, Z)=(0.01,0.03)$, when $t \Omega \sim 100$, $\chi$ rapidly shifts towards $-90^{\circ}$ across the entire $x$-axis and maintains a similar value throughout the simulation time. The rapid change in $\chi$ also corresponds to when fine fluctuations begin to appear in Stokes~$\langle I \rangle_{y}$ in an otherwise smooth start. For $(\St,Z)=(0.03,0.005)$, the situation is fairly different, where $\chi = 0^{\circ}$ for a longer period of time. It takes until $t \Omega \sim 1000$ for the $\langle \epsilon \rangle_{y}$ to increase (Fig.~\ref{fig:spacetime}b), for a large filament to form (Fig.~\ref{fig:spacetime}d) and at the same time, for $\chi$ decrease to values near $-90^{\circ}$ (Fig.~\ref{fig:spacetime}f). The extent in $x$ where $\chi = \pm 90^{\circ}$ corresponds to where $\langle \epsilon \rangle_{y}$ and Stokes~$\langle I \rangle_{y}$ are large (i.e., where dust is concentrated) and moves together with time until the end of the simulation.

Overall, in both simulations, $\chi$ shifts toward $-90^\circ$ as filaments form at the midplane and is maintained until the end of the simulations. The distinguishing behavior between the two cases arises from the different filament evolution pathways (see Sec.~3.2 in \citealt{Lim2026ApJ..1000..156L} for details). For $(\St, Z) = (0.01, 0.03)$, the relatively high $Z$ produces a broad region with an enhanced midplane $\langle \epsilon \rangle_{y}$ from dust settling alone (Fig.~\ref{fig:spacetime}a), resulting in $\chi \sim -90^\circ$ across a wide radial extent early on. In contrast, for $(\St, Z)=(0.03,0.005)$, achieving $\chi = -90^\circ$ further requires a concentration in the radial direction to locally increase the midplane $\langle \epsilon \rangle_{y}$ (Fig.~\ref{fig:spacetime}b) and form the filament occupying only a part of the radial domain.

We now explore all the simulations generated in \cite{Lim2026ApJ..1000..156L} which covers several combinations of $(\St,Z)$. To simplify the exploration, we take only the last snapshot to avoid effects of initial conditions. To be representative of unresolved observations, we calculate $\chi$ utilizing the Stokes~$Q$ and $U$ summed over the spatial domain and search for the parameter space that is better favored by the observed $\chi \sim \pm 90^{\circ}$.
Naturally, spatially averaged $\chi$ cannot capture the complex 2D distribution, but any deviation from $\chi=0^{\circ}$ represents some level of azimuthal flow. As a reference, the spatially averaged $\chi$ is $-76.7^{\circ}$ and $-31.7^{\circ}$ for $(\St,Z)=(0.01,0.03)$ and $(0.03,0.005)$, respectively. The latter demonstrates that even though regions of $\chi = -90^{\circ}$ may exist (Fig.~\ref{fig:pol2d}d), a spatially unresolved $\chi$ may not be near $-90^{\circ}$ if the region is not widespread within the telescope beam.

\begin{figure*}
    \centering
    \includegraphics[width=\textwidth]{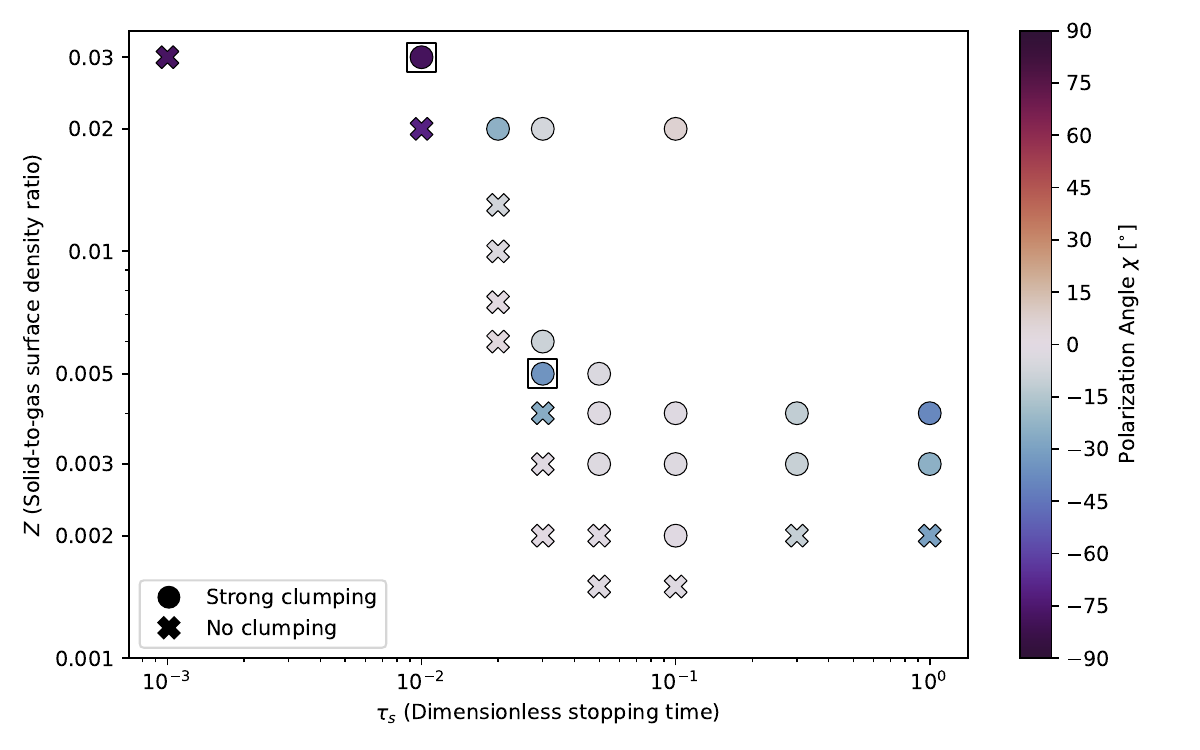}
    \caption{
        The polarization angle $\chi$ using Stokes~$Q$ and $U$ summed over the spatial domain for each simulation taken at the last snapshot. The colors correspond to different levels of $\chi$, where $\chi=0^{\circ}$ is polarization in the radial direction, while $\chi=\pm90^{\circ}$ is polarization in the azimuthal direction. Each simulation is initialized with a different combination of $(\St,Z)$. The circles denote cases with strong clumping, while the crosses denote cases without strong clumping. The squares highlight the two examples explored in more detail in Fig.~\ref{fig:midplane}, \ref{fig:pol2d}, and \ref{fig:spacetime}. 
    }
    \label{fig:pol0d}
\end{figure*}

Fig.~\ref{fig:pol0d} shows the values of $\chi$ for each case. For comparison, we also denote whether the simulations achieved strong clumping, where the maximum density of solid particles exceeds the Hill density, allowing gravitational collapse and planetesimal formation \citep{Li2021ApJ...919..107L, Lim2025ApJ...981..160L}. 
The two simulations with $Z=0.03$ show clear azimuthal $\vect{A}$-flow as well as the case with $(\St,Z)=(0.01, 0.02)$. Beyond these smallest $\St$ cases, the rest do not show equally strong azimuthal flow that dominates the polarized intensity. Thus, it overall appears to be the low $\St$ and high $Z$ regime that are at least more capable of matching observations. However, the detailed dependence on $\St$ and $Z$ is difficult to understand without spatially resolving the high $\epsilon$ regions, which is beyond the scope of this paper. We make note of a few features.

There appears to be a slight preference for larger $Z$ values to show azimuthal polarization. For example, the two highest $Z=0.03$ cases show strong azimuthal $\chi$. For cases $\St=0.02$ and $0.03$, $\chi$ is generally less radial toward higher values of $Z$. Indeed, the previously explored $(\St,Z)=(0.03,0.005)$ case shows clear azimuthal $\chi$ (Fig.~\ref{fig:pol2d}d). However, further increase in $Z$ to 0.02 appears to run opposite to the general trend.

There also appears to be a slight preference for smaller $\St$ values. 
With $Z=0.02$, the small $\St=0.01$ shows clearly $\chi \sim -90^{\circ}$ and the effect decreases when increasing to $\St=0.1$. However, one can also identify that for $Z=0.002$ to 0.004, an azimuthal component reappears when $\St=1$ where $\chi \sim - 40^{\circ}$. Values of $\St=0.1$ appear to strongly inhibit azimuthal $\chi$ and only produce radial $\chi$ regardless of $Z$. 

We note that the suite of simulations used here was originally designed to explore the conditions of strong clumping rather than to target aerodynamic flow geometry. As a result, the parameter coverage was not optimized for analyzing the $\vect{A}$-field. With this indirect design, we reveal that the clear, azimuthal $\chi$ can depend on $\St$ and $Z$. On the other hand, we do not identify any correlation between the azimuthal $\chi$ and the level of clumping. Future simulations customized to study the $\vect{A}$-field and its dependence on $\St$, $Z$, and time in detail would be valuable in understanding the onset of azimuthal flows. We view Fig.~\ref{fig:pol0d} as a starting point for follow-up work.

\section{Discussion} \label{sec:discussion}

\subsection{Departure from the Equilibrium Solution}

\begin{figure*}
    \centering
    \includegraphics[width=\textwidth]{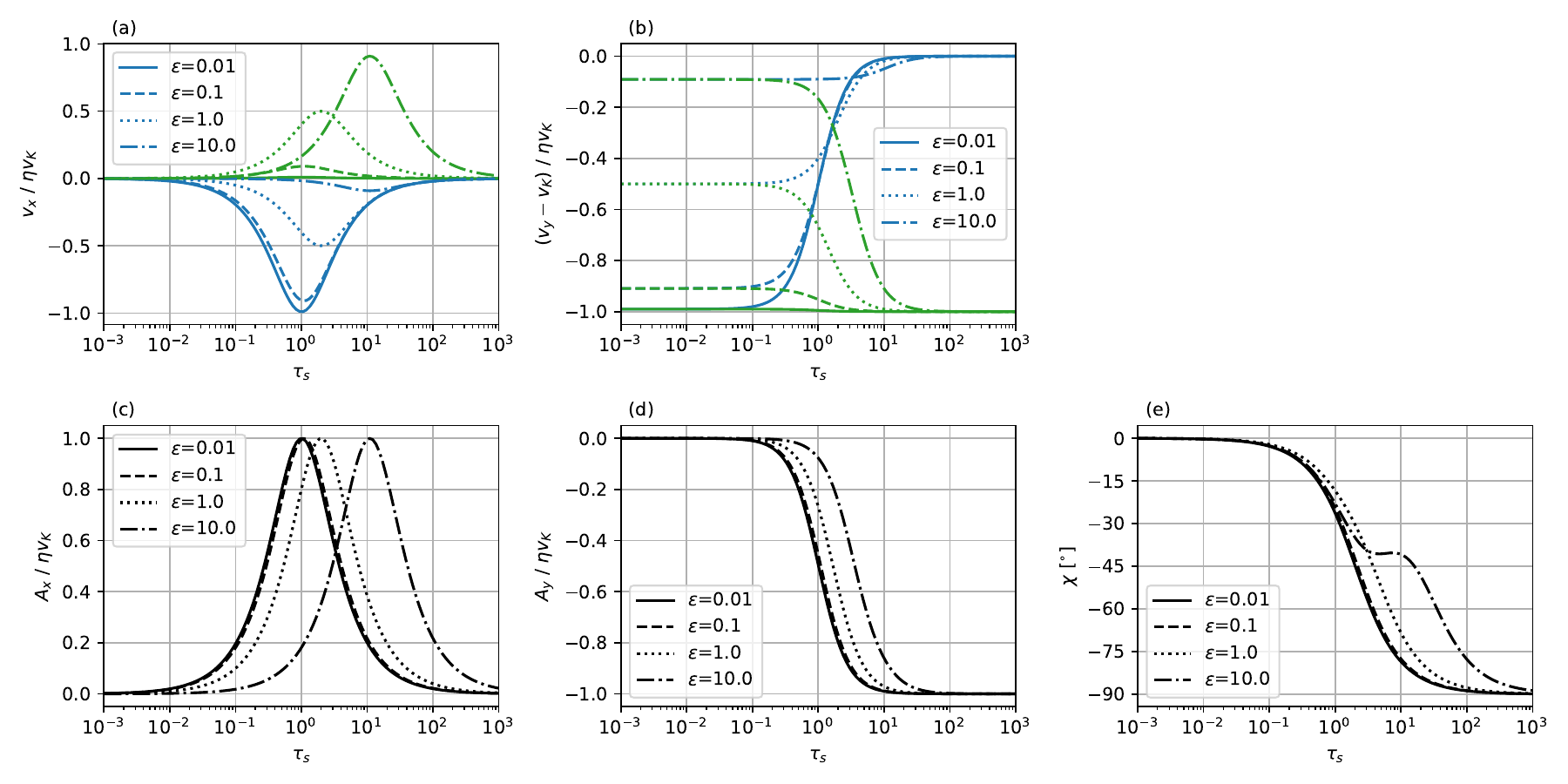}
    \caption{
        The NSH equilibrium solutions of the dust and gas velocity. Panel a and b: the dust (blue) and gas (green) velocity in the radial direction and the azimuthal direction, respectively, as a function of $\St$. The velocity in the azimuthal direction is relative to the Keplerian velocity $v_{K}$. Different line styles correspond to different dust-to-gas ratios $\epsilon$. Panel c and d: the gas velocity relative to the dust in the radial and azimuthal direction, respectively. Panel e: the resulting polarization angle $\chi$ from birdie-like alignment. 
    }
    \label{fig:NSH_solution}
\end{figure*}


Previously, \cite{Lin2024MNRAS.534.3713L} applied birdie-like alignment to a disk velocity field in the limit where there is no dust feedback and found that an azimuthal $\vect{A}$-field requires $\St \gg 1$. Including dust back-reaction, the steady-state solution to the coupled gas and dust dynamics is the Nakagawa—Sekiya—Hayashi (NSH) equilibrium solution \citep{Nakagawa1986Icar...67..375N}.
Following \cite{Birnstiel2024ARA&A..62..157B} (Eq.~10-13), we show the NSH equilibrium solution and the $\vect{A}$-field in Fig.~\ref{fig:NSH_solution}. 
The velocity units are in $\eta v_{K}$ where $v_{K}$ is the Keplerian velocity and 
\begin{align}
    \eta \equiv - \frac{1}{2} \bigg(\frac{H}{r} \bigg)^{2} \frac{\partial \ln P}{\partial \ln r}
    \nonumber
\end{align}
is a factor capturing the pressure gradient, where $P$ is the pressure and $r$ is the radius. The polarization angle $\chi$ is determined from $A_{x}$ and $A_{y}$ directly and only depends on $\St$ and $\epsilon$. 
Fig.~\ref{fig:NSH_solution}e shows that when $\St \ll 1$, the alignment angle is in the radial direction even if we incorporate dust feedback. The alignment angle is in the azimuthal direction only when $\St \gg 1$ regardless of $\epsilon$.

(Sub)millimeter grains in disks are expected to have $\St \sim 10^{-3}$ to $10^{-2}$ \citep{Drazkowska2023ASPC..534..717D}. There have also been millimeter observational constraints utilizing the dust ring widths or level of dust settling which gives $\alpha / \St \sim 0.1$ where $\alpha$ is the typical viscosity parameter \citep[e.g.][]{Dullemond2018ApJ...869L..46D, Villenave2022ApJ...930...11V, Jiang2025ApJ...993..166J, Birnstiel2024ARA&A..62..157B, Villenave2025A&A...697A..64V}. For $\alpha$ at values of $10^{-4}$ to $10^{-2}$, we also obtain $\St \sim 10^{-3}$ to $10^{-1}$. Thus, although $\St \gg 1$ remains one solution simply from a dynamical perspective, it is unlikely the grains that are probed by millimeter wave polarization.

With $\St \ll 1$ and $\chi \sim 0^{\circ}$ (Fig.~\ref{fig:NSH_solution}e) even for high $\epsilon$, dust feedback alone is insufficient to produce the azimuthal $\vect{A}$-field while confined to the axisymmetric and steady-state assumption. \cite{Youdin2005ApJ...620..459Y} found that the system described by the NSH equilibrium is unstable to the growth of perturbations, which is precisely SI. 
This work demonstrates that the azimuthal polarization pattern of $\St \ll 1$ grains requires a departure from the equilibrium solution and specifically SI in the high $\epsilon$ region.

Direct comparisons to observations utilizing SI in global 3D simulations or at least spatial scales comparable to observations are imperative. After all, the high angular resolution observations with the Atacama Large Millimeter/submillimeter Array (ALMA) can resolve the dust rings, but not much more, especially given the high cost of detecting polarization compared to total intensity alone \citep{Andrews2018ApJ...869L..41A, Stephens2023Natur.623..705S}. Since dust rings are likely due to pressure bumps, simulations require considering a more global aspect incorporating changes in the pressure gradient \citep{Carrera2021AJ....161...96C}. While this work has demonstrated azimuthal $\vect{A}$ with a constant pressure gradient, it is unclear how $\vect{A}$ behaves with pressure bumps. 

While this work demonstrates that SI is one dynamical mechanism that can generate the observed azimuthal $\vect{A}$-field, other potential mechanisms exist, including planet-disk interactions, Rossby wave instability, secular gravitational instabilities, or magnetohydrodynamic processes (see \citealt{Bae2023ASPC..534..423B} for a review), but their effects on grain alignment are rarely explored in detail. In one case, \cite{LietzowSinjen2025A&A...703A..60L} studied grain alignment with the Vertical Shear Instability (VSI) and showed that the $\vect{A}$-field is predominantly in the vertical direction. One can thus readily rule out VSI as the dominant mechanism if grains are aligned mechanically. Future works comparing the $\vect{A}$-field from various mechanisms are necessary to verify if SI is a unique solution.

\subsection{The Grain Alignment Problem of Protoplanetary Disks}



As mentioned in Sec.~\ref{sec:intro}, the observed effectively prolate grains have convincingly ruled out the conventional RAT grain alignment paradigm adopted for the ISM. For grains smaller than $\sim 1$~$\mu$m, we generally expect internal energy dissipation, which aligns the short axes of grains to the angular momentum, also called ``internal alignment" \citep{Purcell1979ApJ...231..404P}. How the population aligns as an ensemble depends on the external field. Under strong RAT conditions, the short axes of grains with helicity are expected to be aligned to the direction of radiation anisotropy \citep{Lazarian2007MNRAS.378..910L, Tazaki2017ApJ...839...56T}. Under the MEchanical Torques (MET) framework, the short axes of grains with helicity are aligned to the aerodynamic flow due to gas bombardment \citep{Lazarian2007ApJ...669L..77L, Hoang2018ApJ...852..129H}. In both cases, if there exists a $\vect{B}$-field, the short axes become aligned to the $\vect{B}$-field since spinning, paramagnetic grains gain a magnetic moment through the Barnett effect \citep{Dolginov1976Ap&SS..43..257D, Purcell1979ApJ...231..404P}. 
In any case, the issue lies in the alignment of the short axes, which creates effectively oblate grains. Perhaps it is not too surprising, since in disk environments, where the grains responsible for millimeter polarization are at least 100~$\mu$m and gas densities are drastically higher, internal relaxation is no longer faster than the gas damping time, making internal alignment an invalid assumption \citep{Hoang2009ApJ...697.1316H, Hoang2022AJ....164..248H}.

How large grains can be aligned to external vector fields in the limit of slow internal relaxation is currently unclear.
The birdie-like alignment adopted in this work is one approach utilizing gas bombardment in this limit. 
Experimentally, \cite{Wurm2000ApJ...529L..57W} showed that grains grown in a cluster-cluster type of aggregation are naturally elongated and that they preferentially align their long axes along the flow direction, which is consistent with birdie-like alignment. The long-to-short axes ratio in that direction is $\sim 1.2$ which is intriguingly consistent with that inferred from polarization of HL~Tau \citep{Stephens2023Natur.623..705S, Lin2024MNRAS.528..843L}.

Another candidate is the so-called ``wrong" internal alignment with $\vect{B}$-RAT where the short axis of a grain is perpendicular to its angular momentum $\vect{J}$ and $\vect{J}$ is aligned to $\vect{B}$ at low-$\vect{J}$ attractors in the limit of slow internal relaxation \citep{Hoang2022ApJ...928..102H, Hoang2022AJ....164..248H}. As a result, the polarization is parallel to the $\vect{B}$-field. Under a toroidal $\vect{B}$-field for a disk, the polarization behaves as azimuthally aligned, effectively prolate grains that can satisfy the azimuthal variation in the polarization degree and direction in a disk \citep{Thang2024ApJ...970..114T}. 
While the two candidates cannot be distinguished by polarization morphology, it can be differentiated by the damping time. Short damping time due to high gas densities benefits the birdie-like alignment by stopping all spin, and puts $\vect{B}$-RAT at a disadvantage, since grains need to spin to maintain a magnetic moment from the Barnett effect. 
Several theoretical works have shown that the gas damping time is usually shorter than Larmor precession time in midplanes of disks \citep[e.g.][]{Tazaki2017ApJ...839...56T, Yang2021ApJ...911..125Y, Hoang2022AJ....164..248H}. 

Observationally, \cite{Fourkas2026ApJ...999....4F} demonstrated with Class I BHB 07-11 that the damping timescale tends to be shorter than the Larmor precession time for a wide range in potential grain structure, offering one case that favors birdie-like alignment. 
Intriguingly, \cite{Looney2025ApJ...985..148L} showed that the aligned grains in the Class 0 L1448 IRS3 follows the Stokes~$I$ spiral in the high-density regions, but their directions and polarization fraction change drastically in the low-density, outer regions. The two cases here hint at a gas density dependence where $\vect{B}$-RAT operates in the low gas density regime while birdie-like alignment takes over in the high gas density regime. 
However, in a low-density gas environment such as that of a debris disk, \cite{Hull2022ApJ...930...49H} found that grain alignment is inconsistent with magnetic alignment. Given the upper limit on polarization of approximately $1\%$, any alignment that does occur would be extremely poor, even if $\vect{B}$-RAT were the underlying mechanism.
Several questions about alignment processes remain, and more theoretical clarification is necessary.

At the same time, given the multitude of potential grain structures, observational tests are invaluable. After all, the grain shape and orientation are deduced immediately from the azimuthal variation in the polarization expected from axisymmetric disks and are relatively immune from assumptions of any alignment theory \citep{Yang2019MNRAS.483.2371Y, Mori2021ApJ...908..153M}. Searching for aligned grains and testing their dependence on the environment, for example, the gas density, evolutionary stage, or the $\vect{B}$-field morphology (constrained by other means), can provide tremendous insights.

\subsection{Observational Implications}

The SI has been proposed as the main candidate for planetesimal formation \citep{Chiang2010AREPS..38..493C}, but the observational evidence remains scant for such an important step in planet formation. \cite{Nesvorny2019NatAs...3..808N} identified that the prograde mutual inclinations of binaries in the cold classical Kuiper Belt is consistent with expectations from the SI. The size distribution of the asteroid belt and Kuiper Belt objects can also be compared to the resulting planetesimal population from SI simulations with gravitational collapse of solids \citep[e.g.][]{Simon2016ApJ...822...55S, Li2025ApJ...995..214L}. The density dichotomy of Kuiper Belt objects can also be explained by SI \citep{Canas2024PSJ.....5...55C}. 
The obvious downside of this approach is that the constraints come entirely from the solar system.
As we have demonstrated, birdie-like alignment and the azimuthally aligned prolate grains provide another line of evidence for SI. 
Therefore, we believe that dust polarization opens a new window to directly study the millimeter grains experiencing ongoing SI in protoplanetary systems other than the solar nebula.

HL~Tau is currently the disk with the best set of polarization data, ranging from $\lambda=0.87$~mm to $7$~mm \citep{Stephens2014Natur.514..597S, Kataoka2016ApJ...820...54K, Stephens2017ApJ...851...55S, Lin2024MNRAS.528..843L} along with the highest angular resolution (5~au) at $\lambda=0.87$~mm \citep{Stephens2023Natur.623..705S}. The existence of azimuthally aligned, prolate grains, even within the gaps, has been well established in this Class I/II disk accounting for optical depth effects. 
Several other sources, while studied in less detail, also show the azimuthally oriented polarization pattern that is consistent with azimuthally aligned, prolate grains, including Class I IRS~63 \citep{Sadavoy2019ApJS..245....2S}; other Class II sources, like 
AS~209 \citep{Mori2019ApJ...883...16M, Harrison2021ApJ...908..141H}, Haro 6-13 \citep{Harrison2019ApJ...877L...2H, Harrison2024ApJ...967...40H}, 
DG~Tau \citep{Harrison2019ApJ...877L...2H}, 
V892~Tau \citep{Harrison2024ApJ...967...40H}, 
GG~Tau \citep{Tang2023ApJ...947L...5T}; 
and even a high-mass star GGD~27 MM1 \citep{Girart2018ApJ...856L..27G}. 
Several binaries or triples also show azimuthally oriented polarization patterns, including L1448 IRS3 \citep{Looney2025ApJ...985..148L}, BHB~07-11 \citep{Fourkas2026ApJ...999....4F}, and HD~142527 \citep{Ohashi2025NatAs...9..526O}. 

If the SI paradigm and the birdie-like alignment mechanism are correct, these disks with broad regions of azimuthal polarization pattern require broad regions of SI \textit{and} $\epsilon \gtrsim 1$ in the midplane. Measuring grain alignment in rings versus gaps will be an important comparison. If gaps are truly dust-poor near the midplane, with $\epsilon \ll 1$, and the dust feedback is extremely weak, then we would expect the radial pattern that resembles that from the NSH equilibrium limit (Fig.~\ref{fig:pol2d}b). However, gaps are not guaranteed to have $\epsilon \ll 1$ because of dust settling. Along with simulations that fully consider radially varying the pressure gradients, empirical guidance from high angular resolution observations that can resolve rings and gaps will be valuable.

However, several other sources do not show similar polarization patterns, but only the pattern expected of scattering \citep[e.g.][]{Hull2018ApJ...860...82H, Dent2019MNRAS.482L..29D}. It is unclear whether the contrast is due to a complete lack of SI or the signature of aligned grains is simply hidden by scattering.
Given how we can infer the presense of SI or the magnetic field (or large $\St$ grains in steady state, though unlikely) through polarization, all of which have significant importance in understanding disk evolution and planet formation, further discovery awaits from using instruments sensitive to millimeter-wave polarization, as exemplified by ALMA, the Very Large Array (VLA), and eventually ngVLA. 



\section{Summary} \label{sec:summary}

The grain alignment problem of protoplanetary disks is built on three lines of empirical evidence, namely that the grains are large (at least $100$~$\mu$m in size), effectively prolate, and aligned in the disk azimuthal direction, which conventional alignment theories cannot explain. 
The novel birdie-like alignment allows large grains to be aligned as effectively prolate grains in the direction of the gas velocity relative to the dust velocity ($\vect{A}$-field). 
Through state-of-the-art 3D SI simulations from \cite{Lim2026ApJ..1000..156L}, we find that in regions with high dust-to-gas ratio, dust predominantly experiences azimuthal $\vect{A}$. 
We conduct polarized radiation transfer and find that the polarization angle $\chi$ follows the disk azimuthal direction in the regions of high dust-to-gas ratio. The azimuthal $\chi$ depends on the Stokes number ($\St$) and the initial dust content relative to the gas ($Z$). 
With low dust-to-gas ratios, dust experiences radial $\vect{A}$ resulting in radially oriented polarization.
Therefore, the observed large, azimuthally aligned, prolate grains of protoplanetary disks could be a signature of ongoing SI. 
Should this be true, the significance is twofold. First, it solves the long mysterious origin of aligned grains in protoplanetary disks. Second, it offers new evidence to the long anticipated SI currently favored to form planetesimals. The seemingly different lines of work are natural results of aerodynamic interactions between gas and dust.

\begin{acknowledgments}
We thank the anonymous reviewer for the constructive comments.
We thank Austen Fourkas, Yi-Wei (Jerry) Xu, Chun-Yen Hsu, Sidhant Kumar Suar, Shangjia Zhang, Gerhard Wurm, Joanna Dr{\k{a}}{\.z}kowska, Akimasa Kataoka, Cornelis Dullemond, John Tobin, Adele Plunkett, and Patrick Sheehan for fruitful discussions. J.L. acknowledges support from NASA under the Future Investigators in NASA Earth and Space Science and Technology grant \# 80NSSC22K1322. JBS acknowledges support from NASA through grant 80NSSC25K7398 and from NSF through grant 2407762.
ZYL is supported in part by NASA 80NSSC20K0533, NSF AST-2307199, and the Virginia Institute of Theoretical Astronomy (VITA).
MFL was also partly supported by the Spanish program Unidad de Excelencia María de Maeztu CEX2020-001058-M, financed by MCIN/AEI/10.13039/501100011033, and by the MaX-CSIC Excellence Award MaX4-SOMMA-ICE.
HY is supported in part by the National Natural Science Foundation of China (NSFC) [12473067]. 
\end{acknowledgments}

\begin{contribution}

ZYDL came up with the initial research concept, conducted the polarization analsis, and was responsible for writing the manuscript. 
JL conducted the streaming instability simulations and analysis. 
All authors contributed to the scientific discussion. 


\end{contribution}

%

\software{numpy \citep{harris2020array}, matplotlib \citep{Hunter:2007}
          }


\appendix

\section{Vertical Structure of the $\vect{A}$-field} \label{sec:vertical_structure_of_A}

\begin{figure*}
    \centering
    \includegraphics[width=\columnwidth]{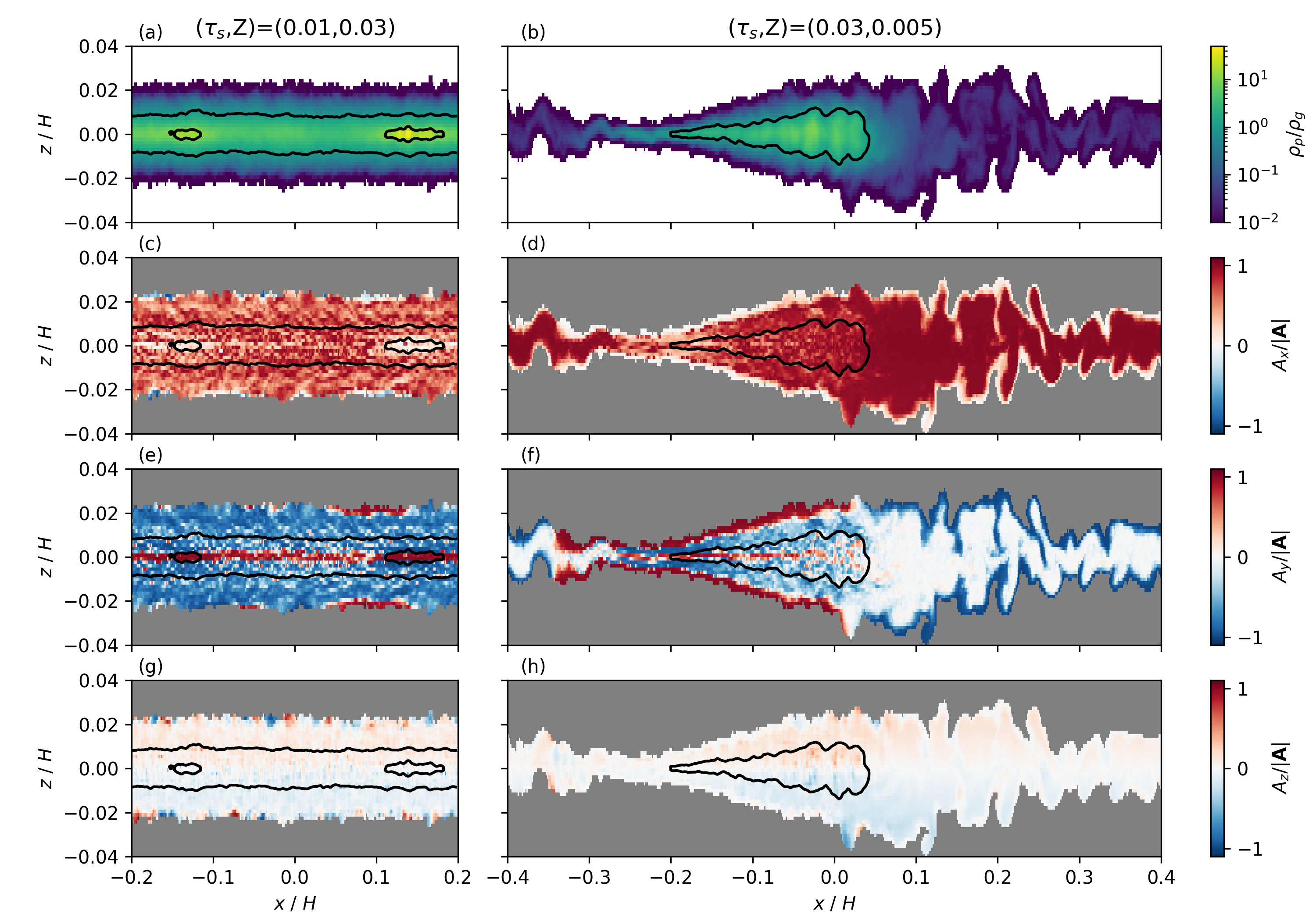}
    \caption{
        The vertical structure of the simulations, where $x$ is the radial direction and $z$ is the vertical direction. The left and right columns correspond to simulations with $(\St,Z)=(0.01,0.03)$ and $(0.03,0.005)$. The top row is the dust-to-gas ratio $\epsilon$. The second, third, and last rows correspond to the radial, azimuthal, and vertical components ($A_{x}$, $A_{y}$, and $A_{z}$) of the aerodynamic flow $\vect{A}$. The gray regions are where $\epsilon < 10^{-5}$.
        The black contours show where $\epsilon=1$ and $10$. Since the dust is highly settled, the horizontal axis covers much more length than the vertical axis for easier visualization. 
    }
    \label{fig:meridian}
\end{figure*}

Here we show the vertical structure of the two example simulations $(\St,Z)=(0.01,0.03)$ and $(0.03, 0.005)$. We average $\epsilon$, $A_{x}$, $A_{y}$, and $A_{z}$ along the $y$-axis for visualization. Fig.~\ref{fig:meridian} shows $\epsilon$ and the normalized components $A_{x}/|\vect{A}|$, $A_{y}/ |\vect{A}|$, and $A_{z} / |\vect{A}|$ to show the relative strength. In both cases, the dust is highly settled in that $\epsilon$ drops below $1$ just beyond $|z|/H \sim 0.01$ (Fig.~\ref{fig:meridian}a,b). For $(\St,Z)=(0.01,0.03)$, $A_{x}/|\vect{A}|$ is generally positive throughout, while $A_{y} / |\vect{A}|$ is positive only in the midplane and turns negative when at regions away from the midplane (Fig.~\ref{fig:meridian}c,e). $A_{z}/ |\vect{A}|$ is positive where $z>0$, but switches sign where $z<0$ giving $A_{z}/|\vect{A}| \sim 0$ as mentioned in Sec.~\ref{sec:results}. It is the smallest among the three components. The $(\St,Z)=(0.03,0.005)$ shows more complex structure. Within $-0.2 < x/H < 0.05$ where the midplane $\epsilon > 1$, $A_{x}$ is generally positive and we see a stratified $A_{y} / |\vect{A}|$ where $A_{y}<0$ in the midplane, but $A_{y}>0$ away from the midplane. The two features are similar to the previous example. However, at $x$ where the midplane $\epsilon<1$, $\vect{A}$ is completely dominated by $A_{x}$, while $A_{y} / |\vect{A}| \sim 0$. The vertical component $A_{z} / |\vect{A}|$ shares a similar change in sign across the midplane like the previous example.

The trailing spiral of polarization $\chi$ is a result of the stratification of $A_{y}$ for both cases. In the midplane where $\epsilon> 1$, $A_{y}>0$ and since $A_{x}$ is also slightly positive, the $\vect{A}$-field in the midplane is a leading spiral. However, above the midplane, where $A_{y}<0$ while $A_{x}>0$, the $\vect{A}$-field is a trailing spiral. Although the midplane has a higher dust density, the layer with $A_{y}>0$ is smaller in its vertical extent compared to the layer where $A_{y}<0$. Polarization from aligned grains in these two layers counter each other, and it is the trailing spiral that wins for these simulations.

\bibliography{main}{}
\bibliographystyle{aasjournalv7}



\end{document}